\documentclass[twocolumn,showpacs,preprintnumbers,amsmath,amssymb,prb]{revtex4}

\usepackage{graphicx}
\usepackage{dcolumn}
\usepackage{bm}
\begin{document}
\newcommand{\ud}{d}


\title{Phase Changes in selected Lennard-Jones 
$\mathrm{X_{13-n}Y_n}$ clusters}

\author{Dubravko Sabo}
\author{Cristian Predescu}
 \altaffiliation[Present address: ]{Department of Chemistry, University of
California, Berkeley, California 94720}
\author{J. D. Doll}
\affiliation{ Department of Chemistry, \\ Brown University, \\ Providence,
\\ Rhode Island, 02912}

\author{David L. Freeman}
\affiliation{Department of Chemistry, \\ University of Rhode Island,
\\ Kingston, Rhode Island, 02881}

\date{\today}

\begin{abstract}
Detailed studies of the thermodynamic properties of selected binary
Lennard-Jones clusters of the type $\mathrm{X_{13-n}Y_n}$ 
(where $\mathrm{n}$=1,2,3) are presented. The total energy, heat capacity and 
first derivative of the heat capacity as a function of temperature are 
calculated by using the classical and path integral Monte Carlo methods 
combined with the parallel tempering technique. A modification in the phase 
change phenomena from the presence of impurity atoms and quantum
effects is investigated.

\end{abstract}


\maketitle

\section{Introduction} \label{sec:intro}

Clusters, as aggregates of atoms or molecules that range in size from
two to tens of thousands of monomer units, can be viewed as an intermediate 
state of matter between finite and bulk. Many of the cluster properties,
such as structural and thermodynamic for example, are different from
the corresponding bulk properties because of the large number of surface
species and finite size effects.

One thermodynamic property of clusters that has received much experimental
\cite{LEUT90,BUCK94C,BUCK02,HABERLAND97,HABERLAND98,HABERLAND99,JARROLD03A} 
and theoretical \cite{BERRY88,BERRY99,CHAKRA95A,CHAKRA95B,CHAKRA03,WALES98A,
CALVO99,CALVO00,CALVO01,JORD93C,JORD98,JORD02A,JORD02B,JORD03A,FRANTZ96,
FRANTZ97,FRANTZ01,FREEMAN92,FREEMAN96,FREEMAN98A,FREEMAN00A,LOPEZ97,LOPEZ02}
attention is the ``phase transition''. Since phase transitions are
characteristic of bulk systems, we shall refer to the phase transformations 
in clusters as phase changes, adopting the language introduced by Berry
and coworkers.\cite{BERRY88}
In a bulk material, the phase transition
from solid to liquid occurs at a definite temperature and the heat
capacity at that temperature has a sharp (delta function like) peak.
\cite{HABERLAND98,HABERLAND99} In clusters, the phase change occurs at
a range of temperatures, lying between freezing and melting temperatures,
and the heat capacity has a broad peak about the transition temperature owing
to finite size effects. Between the freezing and melting temperatures an
ensemble of clusters coexists in liquid-like and solid-like forms.
\cite{BERRY99}
With some notable exceptions\cite{JARROLD03A}, 
the melting transition takes place at lower temperatures than the 
corresponding bulk.\cite{HABERLAND98,HABERLAND99}

In previous work\cite{SABO04A} (hereafter referred to as paper 1), we 
have explored the energy landscape of the selected binary Lennard-Jones 
clusters of the type $\mathrm{X_{13-n}Y_n}$ and examined the effect of 
adding $\mathrm{n}$ $\mathrm{Y}$ impurity atoms on the structures of their 
$\mathrm{X_{13-n}}$ core. In the present work, we extend our studies of the 
given systems into the thermodynamic domain. Thermodynamic properties, such as
the total energy, heat capacity and first derivative of the heat
capacity as a function of temperature, 
are calculated in both the classical and quantum regime.
Our goals are to examine the effect of perturbing atoms on the thermodynamic
properties of the selected systems and to understand how the complex
topology of the potential energy surface affects phase change phenomena.
In addition, we examine the importance of quantum contributions to the
phase change phenomena in these systems.

From a computational point of view, these clusters are also
interesting because of their complex potential energy surface. 
The double-funnel
character of their potential energy surface makes them a particularly
challenging case for Monte Carlo simulations.\cite{FREEMAN00A,FREEMAN00B,
JORD02A} Therefore, they constitute a good numerical test for measuring
the efficiency of Monte Carlo techniques designed to overcome
quasiergodicity in both classical\cite{FREEMAN00A,FREEMAN00B} and
quantum\cite{PREDESCU03F} simulations.

The remainder of the paper is organized as follows. In Sec.~\ref{sec:comput}
we give a brief review of the methods and the model potential we employ 
to calculate the thermodynamic properties for the given systems. 
In Sec.~\ref{sec:res} we present the results including the total energy, the 
heat capacity and the first derivative of the heat capacity as a function of 
temperature. The phase change behavior is characterized with the help of the 
probability distribution of isomers as a function of energy.
Finally, in Sec.~\ref{sec:conclude} we summarize our findings.

\section{Computational Method} \label{sec:comput}

In the present Section, we describe the computational details of our
studies involving binary clusters of the form $\mathrm{X_{13-n}Y_n}$. 
We have chosen to study three systems: $\mathrm{X_{12}Y_1}$, 
$\mathrm{X_{11}Y_2}$, and $\mathrm{X_{10}Y_3}$. 
The choice of the systems is motivated by the
detailed studies of their PES and construction of the associated
disconnectivity graphs in the companion paper.\cite{SABO04A} 

\subsection{Interaction potential}
The clusters are modeled by the total potential energy
\begin{equation}
\label{2.1} 
V_{tot} = \sum_{i<j}^{N} V_{LJ}(r_{ij})+\sum_{i=1}^{N}
V_c(\mathbf{r_i}),
\end{equation} 
where $V_{LJ}(r_{ij})$, the pairwise Lennard-Jones potential as
a function of the distance $r_{ij}$ between atoms $i$ and $j$, is given by
\begin{equation}
\label{2.2} 
V_{LJ}(r_{ij}) = 4\epsilon_{ij}\left
        [\left( \frac{\sigma_{ij}}{r_{ij}}\right)^{12}
       -\left( \frac{\sigma_{ij}}{r_{ij}}\right)^{6}\right],
\end{equation} 
and V$_{c}(\mathbf{r_i})$ is the confining potential
\begin{equation}
\label{2.3}
V_c(\mathbf{r_i})=\epsilon
\left(\frac{|\mathbf{r_i}-\mathbf{R_{cm}}|}{R_c}\right)^{20}.
\end{equation}
In Eq. (\ref{2.2}), the constants $\epsilon_{ij}$ and $\sigma_{ij}$ are
the energy and length-scale parameters for the interaction of
particles $i$ and $j$.

For the binary clusters, we need to specify both the ``like'' 
($\mathrm{X}$-$\mathrm{X}$, $\mathrm{Y}$-$\mathrm{Y}$) as well as
the ``mixed'' ($\mathrm{X}$-$\mathrm{Y}$) interactions. 
The ``mixed'' Lennard-Jones
parameters are obtained from the ``like'' Lennard-Jones parameters by
usual combination rules \cite{COMBINE}
\begin{equation}
\label{2.4}
\sigma_{XY}= \frac{1}{2}(\sigma_{XX}+\sigma_{YY}),
\end{equation}
\begin{equation}
\label{2.5}
\epsilon_{XY}= \sqrt{\epsilon_{XX}\epsilon_{YY}} .
\end{equation}
In the present work, we have chosen to examine selected binary Lennard-Jones
clusters of the type $\mathrm{X_{13-n}Y_n}$ where the impurity atoms are
less massive than their core ($\mathrm{X}$) counterparts. 
Based on our previous work, \cite{SABO04A} we have chosen
the $\mathrm{X}$-atoms to be argon ($\epsilon_{XX}=119.8$~K,
$\sigma_{XX}=3.405$ {\AA}, $\mathrm{mass}=39.948$) and $\mathrm{Y}$-atoms 
to be ``neon-like''. Specifically, we have chosen the 
($\sigma= \sigma_{YY}/\sigma_{XX}$, $\epsilon= \epsilon_{YY}/\epsilon_{XX}$)
ratios for the impurity atoms to be (0.8,0.6) for $\mathrm{X_{12}Y_1}$ and
(0.8,0.5) for both $\mathrm{X_{11}Y_2}$ and $\mathrm{X_{10}Y_3}$, values
that produce interesting classes of potential energy surfaces topologies.
The mass of the $\mathrm{Y}$-atom impurity has been taken to be that of 
of neon (20.1797) in all calculations.

In Eq. (\ref{2.3}) $\mathbf{r_{i}}$ and $\mathbf{R_{cm}}$ are the
coordinates of the $i$th atom and the center of mass of the cluster,
respectively. The center of mass of the cluster is given by
\begin{equation}
\label{2.6}
\mathbf{R_{cm}}=\frac{\sum_{i=1}^{N} m_i \mathbf{r_i}}
{\sum_{i=1}^{N} m_i}.
\end{equation}
Finally, $R_c$ is the radius of the confining sphere\cite{ABRAHAM73}
while $\epsilon$ governs the strength of the confining potential. 
The role of the confining potential $V_c(\mathbf{r_i})$ is to prevent
atoms from permanently leaving the cluster since the cluster in
vacuum at any finite temperature is metastable with respect to
evaporation.

The optimal choice of the parameter $R_c$ for the confining potential
has been discussed in recent work.\cite{NEIROTTI00}  If $R_c$ is taken to be
too small, the properties of the system become sensitive to its choice,
whereas large values of $R_c$ can result in problems attaining an ergodic
simulation. The classical and quantum Monte Carlo simulations presented here
have been performed with $R_c=2\sigma_{XX}$ and $\epsilon=\epsilon_{XX}$.
Since the potential energy surfaces of all the systems studied in the
present work display a double-funnel character,\cite{SABO04A} their
thermodynamic properties are calculated with Monte Carlo
methods coupled with the parallel tempering technique \cite{PARISI92,
THOMPSON95,HANSMANN97,DEEM99A,DEEM99B,PABLO99,FREEMAN00A,FREEMAN00B,
PREDESCU03F}
devised to tackle possible ergodicity problems.

\subsection{Classical Monte Carlo simulation}

The total energy $\langle E \rangle$, the constant volume heat capacity
$\langle C_V \rangle$, and the first derivative of the heat capacity with 
respect to the temperature ($\partial \langle C_V \rangle/\partial T$)$_V$
for clusters consisting of $N$ particles 
are calculated using the standard expressions
\begin{equation}
\label{2.7}
\left\langle E \right\rangle=\frac{3}{2\beta}N + \left\langle V \right\rangle,
\end{equation}
\begin{equation}
\label{2.8}
\frac{\langle C_V \rangle}{k_B}=
\frac{3}{2}N + \beta^2[\langle V^2 \rangle-\langle V \rangle^2],
\end{equation}
\begin{eqnarray} \nonumber
\label{2.9} &&
\frac{1}{k_B}\left(\frac{\partial \langle C_V \rangle}{\partial T}\right)_V \\
 && \quad=
-2\beta \langle C_V \rangle + \beta^4
[\langle V^3 \rangle+2\langle V \rangle^3
-3\langle V^2 \rangle\langle V \rangle], \qquad
\end{eqnarray}
where $k_B$ is Boltzmann constant, $T$ is the temperature, $\beta=1/k_BT$,
$V$ is the potential energy and angular brackets denote the canonical
averages with respect to the Boltzmann weight exp(-$\beta V(\mathbf{x}))$.

\subsubsection{Parallel tempering and sampling strategy}

The parallel tempering Monte Carlo simulations are carried out using a
total of 48 parallel streams, each running a replica of the system at a
different temperature. The streams are independent and uncorrelated sequences
of random numbers that can be generated simultaneously on
multiple processors. In this paper,
we have used the parallel random number generator library called
Scalable Parallel Random Number Generator (SPRNG).\cite{CEPERLEY99,CEPSPRNG}
Temperatures are generated in the range from $T_{min}$ to $T_{max}$ in such 
a way that they are distributed by geometric progression\cite{PREDESCU04A}
\begin{equation}
\label{2.10}
T_j=R_T^{j-i} T_{min} \quad \ 1\leq j \leq M,
\end{equation}
where
\begin{equation}
\label{2.11}
R_T=(T_{min}/T_{max})^{1/(M-1)} .
\end{equation}
We have chosen $T_{min}=0.2$ K, $T_{max}=50$ K and $M=48$. 
For the range of temperatures $[T_{min},T_{max}]=[0.2, 50]$~K, 
a number of $M = 48$ streams has produced acceptance probabilities for swaps
larger than $40\%$ for all streams and for all simulations performed.

Explicitly, the Monte Carlo simulation is conducted as follows. 
For each stream,
a random walk is carried out through configuration space according to
the Metropolis algorithm.\cite{KALOS86} The basic Monte Carlo step consists
of attempted moves of the physical coordinates associated with a given
particle. Each attempted move is either accepted or rejected in accord
with the Metropolis prescription. The maximum displacements 
have been adjusted in order
to ensure a 40\%-60\% acceptance ratio in the Monte Carlo moves. The 
maximum displacements or step sizes are chosen in an analogous manner
to the temperatures, to satisfy geometric progression.
In other words, the step size at the
temperature $T_j$ is 
\begin{equation}
\label{2.12}
s_{\alpha j}=R^{1-j} s_{\alpha,min},
\end{equation}
where
\begin{equation}
\label{2.13}
R=(s_{\alpha,min}/s_{\alpha,max})^{1/(M-1)} \quad (\alpha=x,y,z). 
\end{equation}
The required acceptance ratio between 40\% and 60\% is achieved for
$s_{\alpha,min}=0.3$ au, $s_{\alpha,max}=1.2$ au and $M=48$.

We define a \emph{pass} as the minimal set of Monte Carlo attempted moves
over all particles in the system. Since the clusters of interest are
made of 13 atoms, a pass consists of 13 basic steps. We also define a
\emph{block} as a set of ten thousand passes. The size of the block is
sufficiently large that the block averages of the estimated
quantities are independent for all practical purposes. The simulation
is divided in two phases: an equilibration phase that consists of 100
blocks and an accumulation phase that consists of 400 blocks per temperature.

An exchange of configurations between streams at adjacent temperatures
has been attempted every 25 passes and it has been accepted or rejected
according to the parallel tempering logic.\cite{PARISI92,THOMPSON95, 
HANSMANN97,DEEM99A,DEEM99B,PABLO99,FREEMAN00A,FREEMAN00B,PREDESCU03F}
A stream at any given
temperature attempts a swap of configurations
with a stream at adjacent lower and higher temperature in succession.
Because of this swapping strategy, the streams at minimum and maximum
temperatures are involved in swaps only every 50 passes. 

All error bars quoted in the current work correspond to two 
standard deviations.
In order to avoid the cluttering of data the error bars have not been
shown. The error bars are comparable to the thickness of the lines drawn
in the various graphs. \\ 

\subsection{Path integral Monte Carlo simulation}

For quantum simulations of the heat capacity, we have employed
a reweighted Wiener-Fourier path integral (RW-WFPI) method 
\cite{PREDESCU03B,PREDESCU03C} and recently developed
energy and heat capacity estimators that can be numerically implemented
by finite difference schemes.\cite{PREDESCU02,PREDESCU03F}
Since methodology is fully described in the cited references 
in this section we only present their salient features.

The quantum analogs of the total energy $\langle E \rangle$, 
the constant volume heat capacity
$\langle C_V \rangle$, and its first derivative with
respect to the temperature ($\partial \langle C_V \rangle/\partial T$)$_V$
are given with the following expressions
\begin{equation}
\label{2.14}
\left\langle E \right\rangle=-\frac{1}{Z}
\left(\frac{\partial Z}{\partial\beta}\right)_V,
\end{equation}
\begin{equation}
\label{2.15}
\frac{\langle C_V \rangle}{k_B}=
\frac{\beta^2}{Z} \left(\frac{\partial^2Z}{\partial\beta^2}\right)_V
-\left[\frac{\beta}{Z} 
\left(\frac{\partial Z}{\partial\beta}\right)_V\right]^2,
\end{equation}
\begin{widetext}
\begin{eqnarray}
\label{2.16}
\frac{1}{k_B}\left(\frac{\partial \langle C_V \rangle}{\partial T}\right)_V=
-2\beta \langle C_V \rangle  -k_B\beta \left\{
\frac{\beta^3}{Z}\left(\frac{\partial^3Z}{\partial\beta^3}\right)_V
+2\left[\frac{\beta}{Z}\left(\frac{\partial Z}{\partial\beta}\right)_V\right]^3 
-3\left[\frac{\beta^2}{Z}\left(\frac{\partial^2Z}{\partial\beta^2}\right)_V
\right]
\left[\frac{\beta}{Z}\left(\frac{\partial Z}{\partial\beta}\right)_V\right] 
\right\},
\end{eqnarray}
\end{widetext}
where $Z$ is a partition function of the system. The partition function
is an integral of the diagonal density matrix over whole configuration
space
\begin{equation}
\label{2.17}
Z = \int_{\mathbb{R}^{3N}} \rho(\mathbf{x};\beta)d\mathbf{x}~.
\end{equation}
In the random series path-integral representation,\cite{PREDESCU02} the
density matrix can be written
as follows 
\begin{eqnarray}
\label{2.18}
\frac{\rho(\mathbf{x};\beta)}{\rho_{fp}(\beta)}&
=&\int_{\Omega^{3N}}\ud P[\bar{\mathbf{a}}]\nonumber
\exp\bigg\{-\beta \int_{0}^{1}\! \!
V\Big[\mathbf{x} \\& +& \sigma \sum_{k=1}^{\infty}\mathbf{a}_k
\Lambda_k(u) \Big]\ud u\bigg\},
\end{eqnarray}
where $\rho_{fp}(\beta)$ is the density matrix of a free particle
[for the $i$th free particle $\rho_{fp,i}(\beta)=
(m_i/2\pi\hbar^2\beta)^{1/2}$ ],
$\ud P[\bar{\mathbf{a}}]$ is the probability measure, defined on the
space $\Omega^{3N}$
\begin{equation}
\label{2.19}
\ud P[\bar{\mathbf{a}}]=\prod_{i=1}^{N}\ud P[\bar{a}_i],
\end{equation}
with
\begin{equation}
\label{2.20}
\ud P[\bar{a}_i] = \prod_{k=1}^{\infty}\ud \mathbf{a_{i,k}}
\frac{1}{\sqrt{2\pi}} e^{-\mathbf{a_{i,k}}^2/2}~.
\end{equation}
The vectors $\mathbf{a}_k^T=(\mathbf{a_1}_{,k},\mathbf{a_2}_{,k},\ldots,
\mathbf{a_N}_{,k})$ are path variables or independent identically distributed
(i.i.d) standard normal variables with $\mathbf{a_i}_{,k}=
(a_{xi,k},a_{yi,k},a_{zi,k})$. The vector $\mathbf{x}^T=(\mathbf{x_1},
\mathbf{x_2},\ldots,\mathbf{x}_{N})$ represents the physical variables
of the system, while $\sigma$ for the $i$th particle is 
$(\hbar^2\beta / m_{i})^{1/2}$ where $m_i$ is its mass. $\sigma \mathbf{a}_k$
can be written as
\[\sigma \mathbf{a}_k = \left(\begin{array}{c}\sigma_1 a_{1,k}\\ 
\vdots \\ \sigma_N a_{N,k}\end{array}\right).\]
The reader should note that $\sigma_i=(\sigma_{xi},\sigma_{yi},\sigma_{zi})$
and $\sigma_{xi}=\sigma_{yi}=\sigma_{zi}$. $\Lambda_k(u)$ are a set of
functions defined in the following way. Let $\{\lambda_k(\tau)\}_{k \geq 1}$
be a set of functions defined on the interval $[0,1]$ that, together
with the constant function $\lambda_0(\tau)=1$, make up an orthonormal
basis in $L^2[0,1]$. Then, $\Lambda_k(u)$ is defined as
\[ \Lambda_k(u)=\int_0^u \lambda_k(t)\ud t.\]

In practical applications, the series in Eq.~(\ref{2.18}) needs to be 
replaced by a finite sum. Within the RW-WFPI framework, a finite
dimensional approximation to the exact density matrix in
Eq.~(\ref{2.18}) is given by the expression
\begin{eqnarray}
\label{2.21}&&
\frac{\rho_n(\mathbf{x};\beta)}{\rho_{fp}(\beta)}
=\int_{\Omega^{3N}} dP[\bar{\mathbf{a}}]\nonumber  \\&& 
\times \exp\bigg\{-\beta \; \int_{0}^{1}\! \!
V\Big[\mathbf{x}+ \sigma \sum_{k=1}^{4n}\mathbf{a}_k
\tilde{\Lambda}_{n,k}(u)\Big]\ud u\bigg\}.\qquad
\end{eqnarray}
The functions,
\begin{equation}
\label{2.22}
\tilde{\Lambda}_{n,k}(u)=
\sqrt{\frac{2}{\pi^2}}  \frac{\sin(k \pi u)}{k}
\end{equation}
for $1 \leq k \leq n$ and
\begin{equation}
\label{2.23}
\tilde{\Lambda}_{n,k}(u)=
f(u)\sin(k\pi u)
\end{equation}
for  $n<k \leq 4n$, are chosen so that to maximize the rate of convergence,
\[
\rho_n(\mathbf{x};\beta) \to \rho(\mathbf{x};\beta).\]
The function $f(u)$ is defined as\cite{PREDESCU03B}
\begin{equation}
\label{2.24} \nonumber
f(u)=\left[\frac{u(1-u)-2/\pi^2\sum_{k=1}^n \sin^2(k\pi u)/k^2}
{\sum_{k=n+1}^{4n}\sin^2(k\pi u)}\right]^{1/2}.
\end{equation}
Therefore, a path that starts and ends in the same configuration space
$\mathbf{x}$ (called a thermal loop) can be written
\begin{eqnarray}
\label{2.25} \nonumber &&
\mathbf{x}(u)=\mathbf{x} +~\sigma\Bigg[\sum_{k=1}^n~a_k~ 
\sqrt{\frac{2}{\pi^2}} \frac{\sin(k\pi u)}{k} \nonumber \\&&
+f(u)\sum_{k=n+1}^{4n}~a_k~\sin(k\pi u)\Bigg]. \nonumber
\end{eqnarray}
It turns out to be useful to define several auxiliary quantities
\cite{PREDESCU03C}
\begin{equation}
\label{2.26} 
U_n(\mathbf{x},\bar{\mathbf{a}};\beta)
=\int_{0}^{1}\! \! V\left[\mathbf{x}(u)\right]\ud u,
\end{equation} 
\begin{equation}
\label{2.27}
X_n(\mathbf{x},\bar{\mathbf{a}};\beta)=
\rho_{fp}(\beta)
 \exp\left[-\beta U_n(\mathbf{x},\bar{\mathbf{a}};\beta)\right],
\end{equation}
and
\begin{eqnarray}
\label{2.28} &&
\nonumber  R_n(\mathbf{x},\bar{\mathbf{a}};\beta,\epsilon)= 
\frac{X_n(\mathbf{x},\bar{\mathbf{a}};\beta \epsilon)}
{X_n(\mathbf{x},\bar{\mathbf{a}};\beta)} =
\epsilon^{-3N/2}\\ && 
\times \exp\left[-\beta\epsilon U_n(\mathbf{x},\bar{\mathbf{a}};\beta\epsilon)
+ \beta U_n(\mathbf{x},\bar{\mathbf{a}};\beta)\right].
\end{eqnarray}
Then it is easy to show that, for $k = 1, 2, 3$, we have
\begin{eqnarray}
\label{2.29} &&
\frac{\beta^k}{Z}\left(\frac{\partial^k Z}{\partial\beta^k}\right)_V =  
\nonumber \\ &&
 \frac{\int_{\mathbb{R}^{3N}}\ud \mathbf{x}\int_{\Omega^{3N}}\ud
P[\bar{\mathbf{a}}]X_n(\mathbf{x},\bar{\mathbf{a}};\beta) 
\frac{d^k}{d\epsilon^k} R_n(\mathbf{x},\bar{\mathbf{a}};\beta,\epsilon)
\Big|_{\epsilon = 1} }
{\int_{\mathbb{R}^{3N}}\ud \mathbf{x}\int_{\Omega^{3N}}\ud
P[\bar{\mathbf{a}}]X_n(\mathbf{x},\bar{\mathbf{a}};\beta)}. \qquad
\end{eqnarray}
The quantities above can be evaluated by Monte Carlo integration,
where the canonical averages are carried out with respect to the 
Boltzmann-like weight $\exp[-\sum_{i=1}^N\sum_{k=1}^{4n}\mathbf{a_{i,k}}^2/2
-\beta U_n(\mathbf{x},\bar{\mathbf{a}};\beta)]$.
The derivatives of the quantity $R_n(\mathbf{x},\bar{\mathbf{a}};\beta)$
can be expressed in terms of the derivatives of the quantity
$U_n(\mathbf{x},\bar{\mathbf{a}};\beta)$
\begin{eqnarray}
\label{2.32}\nonumber
\frac{d}{d\epsilon} R_n(\mathbf{x},\bar{\mathbf{a}};\beta,\epsilon)
\Big|_{\epsilon = 1} 
 &=&  -\frac{3N}{2} - \beta U_n(\mathbf{x},\bar{\mathbf{a}};\beta) \\ &- &
\beta \frac{d}{d\epsilon} U_n(\mathbf{x},\bar{\mathbf{a}};\beta\epsilon)
\Big|_{\epsilon = 1}, \qquad
\end{eqnarray}
\begin{eqnarray}
\label{2.33}\nonumber
\frac{d^2}{d\epsilon^2} R_n(\mathbf{x},\bar{\mathbf{a}};\beta,\epsilon)
\Big|_{\epsilon = 1}  
 &=& \left[\frac{d}{d\epsilon} R_n(\mathbf{x},\bar{\mathbf{a}};\beta,\epsilon)
\Big|_{\epsilon = 1}\right]^2 \\ &+ & \frac{3N}{2} - 2\beta \frac{d}{d\epsilon}
 U_n(\mathbf{x},\bar{\mathbf{a}};\beta\epsilon)\Big|_{\epsilon = 1} 
\nonumber \\ &- &
 \beta \frac{d^2}{d\epsilon^2} U_n(\mathbf{x},\bar{\mathbf{a}};\beta\epsilon)
\Big|_{\epsilon = 1} \qquad
\end{eqnarray}
and
\begin{eqnarray}
\label{2.34}\nonumber
\frac{d^3}{d\epsilon^3} R_n(\mathbf{x},\bar{\mathbf{a}};\beta,\epsilon)
\Big|_{\epsilon = 1}&=&
\left[\frac{d}{d\epsilon} R_n(\mathbf{x},\bar{\mathbf{a}};\beta,\epsilon)
\Big|_{\epsilon = 1}\right]^3 \\ &
+ &3\left[\frac{d}{d\epsilon} R_n(\mathbf{x},\bar{\mathbf{a}};\beta,\epsilon)
\Big|_{\epsilon = 1}\right] \nonumber \\ &
\times & \Big[\frac{3N}{2} - 2\beta \frac{d}{d\epsilon}
 U_n(\mathbf{x},\bar{\mathbf{a}};\beta\epsilon)\Big|_{\epsilon = 1}
\nonumber \\ &
- &\beta \frac{d^2}{d\epsilon^2} U_n(\mathbf{x},\bar{\mathbf{a}};\beta\epsilon)
\Big|_{\epsilon = 1}\Big] \nonumber \\ &
+ &\Big[\frac{3N}{2} - 2\beta \frac{d}{d\epsilon}
 U_n(\mathbf{x},\bar{\mathbf{a}};\beta\epsilon)\Big|_{\epsilon = 1}
\nonumber \\ &
- &\beta \frac{d^2}{d\epsilon^2} U_n(\mathbf{x},\bar{\mathbf{a}};\beta\epsilon)
\Big|_{\epsilon = 1}\Big] \nonumber \\ &
-&\Big[3N + 3\beta \frac{d^2}{d\epsilon^2}
U_n(\mathbf{x},\bar{\mathbf{a}};\beta\epsilon)\Big|_{\epsilon = 1} 
\nonumber \\ &
+ &\beta \frac{d^3}{d\epsilon^3} U_n(\mathbf{x},\bar{\mathbf{a}};\beta\epsilon)
\Big|_{\epsilon = 1}\Big].
\end{eqnarray}
The derivatives with respect to $\epsilon$ are evaluated numerically
by a finite difference approximation
\begin{eqnarray*}&& 
\frac{d}{d\epsilon} 
U_n[\mathbf{x},\bar{\mathbf{a}};\beta \epsilon]\Big|_{\epsilon = 1}  
\approx (2\epsilon_0)^{-1} \\ && 
\times \{U_n[\mathbf{x},\bar{\mathbf{a}};\beta (1+\epsilon_0)]
- U_n[\mathbf{x},\bar{\mathbf{a}};\beta (1-\epsilon_0)]\},
\end{eqnarray*}
\begin{eqnarray*} \\ && 
\frac{d^2}{d\epsilon^2} 
U_n(\mathbf{x},\bar{\mathbf{a}};\beta \epsilon)\Big|_{\epsilon = 1} 
\approx \epsilon_0^{-2} 
\{U_n[\mathbf{x},\bar{\mathbf{a}};\beta(1+\epsilon_0)] \\ &&
- 2 U_n[\mathbf{x},\bar{\mathbf{a}};\beta] 
+   U_n[\mathbf{x},\bar{\mathbf{a}};\beta(1-\epsilon_0)]\},
\end{eqnarray*} 
and
\begin{eqnarray*}&&
\frac{d^3}{d\epsilon^3}
U_n(\mathbf{x},\bar{\mathbf{a}};\beta \epsilon)\Big|_{\epsilon = 1}
\approx \epsilon_0^{-3}
\{  U_n[\mathbf{x},\bar{\mathbf{a}};\beta(1+2\epsilon_0)] \\&&
- 3 U_n[\mathbf{x},\bar{\mathbf{a}};\beta(1+\epsilon_0)]
+ 3 U_n[\mathbf{x},\bar{\mathbf{a}};\beta]
-   U_n[\mathbf{x},\bar{\mathbf{a}};\beta(1-\epsilon_0)]\}.
\end{eqnarray*}

\subsubsection{Parallel tempering and sampling strategy}

The sampling strategy is similar to the one employed in the classical
Monte Carlo simulations. 
Here, with each attempted move of the physical coordinate $\textbf{x}_i$ of 
a particle, we attempt to move a randomly selected path variable associated
with that particle.
A \emph{pass} is defined as the minimal set of Monte Carlo attempted
moves over all 13 particles in the system. We define a \emph{block} as a
set of ten thousand passes. Exceptions are simulations with 32 and 64
path variables, where a block contains twenty thousand passes.
For each simulation 400 blocks have been utilized.

The maximum displacements
for the physical coordinates are chosen analogously to those
utilized in classical simulations [see Eqs.(\ref{2.12}-\ref{2.13})], 
while the maximum displacements for the path variables 
are chosen in the following way: $\Delta_{kj}=As_{xj}$ at temperature $T_j$,
where $A$ is a constant chosen so
that the acceptance ratio for each randomly selected path variable is
between 40\% and 60\%.

We implement a parallel tempering procedure that is analogous to the
one utilized in classical simulations.
An exchange of configurations between streams at adjacent temperatures
has been attempted every 25 passes and it has been accepted or rejected
according to the parallel tempering prescription.
By monitoring the acceptance ratios for all 48 streams we have found that
values have been larger than 40\% for all simulations performed.
It should be noted that the exchange of configurations between streams
includes both particle (physical)
coordinates as well as their associated path variables.

The value of discretization step $\epsilon_0$ needed to evaluate
a finite difference approximation to the derivatives with respect to
$\epsilon$ has been set to $\epsilon_0$=2$^{-12}$. The order of error
for the third derivative is $O(\epsilon_0)$ while for the first and
the second derivative is $O(\epsilon_0^2)$. We would like to
mention that the error introduced in the evaluation of the first
derivative of the heat capacity (total energy, heat capacity), from
the finite difference approximation, is at least ten (one thousand)
times smaller than the corresponding statistical error
for all simulations performed.

We end this section with a comment on the convergence and the error bars
involved in the determination of the total energy and the heat capacity. 
The convergence of the total energy and the heat capacity has been tested
with respect to the number of path variables. We have found the results
to be converged for $4n=32$ path variables. Further increasing of the
number of path variables to 64 has yielded the same results within the
statistical error (two standard deviations).
The error bars have not been shown in the graphs in order 
to avoid the cluttering
of data. For all but the lowest temperatures the error bars are comparable
to the thickness of the lines displayed in the graphs. For reference, at
the lowest temperature at which the path integral simulations  
have been performed,
$T=4$ K, the following results have been obtained for $\mathrm{X_{10}Y_3}$:
$\langle E \rangle=-296.47 \pm 0.13$, $\langle C_V \rangle/Nk_B=0.52 \pm 0.20$,
$(\partial \langle C_V \rangle/\partial T)_V/Nk_B=-0.10 \pm 0.54$.
For the same system but at the temperature 7.22 K we have obtained:
$\langle E \rangle=-293.74 \pm 0.07$, $\langle C_V \rangle/Nk_B=1.10 \pm 0.04$,
$(\partial \langle C_V \rangle/\partial T)_V/Nk_B=0.15 \pm 0.06$.
\begin{figure*}
  \begin{tabular}{@{}cc@{}}
    \includegraphics[width=8.5cm,clip=true]{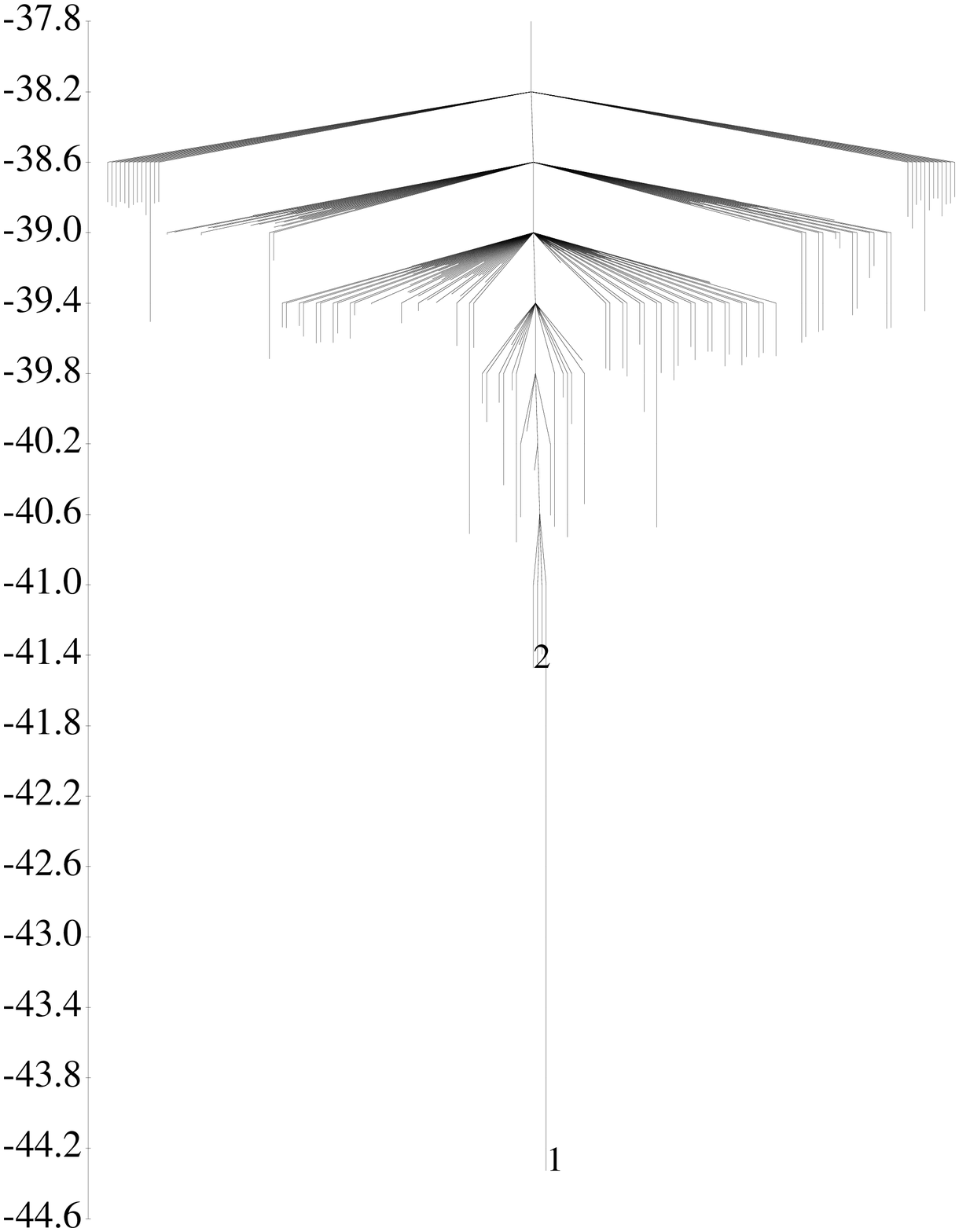} &
    \includegraphics[width=8.5cm,clip=true]{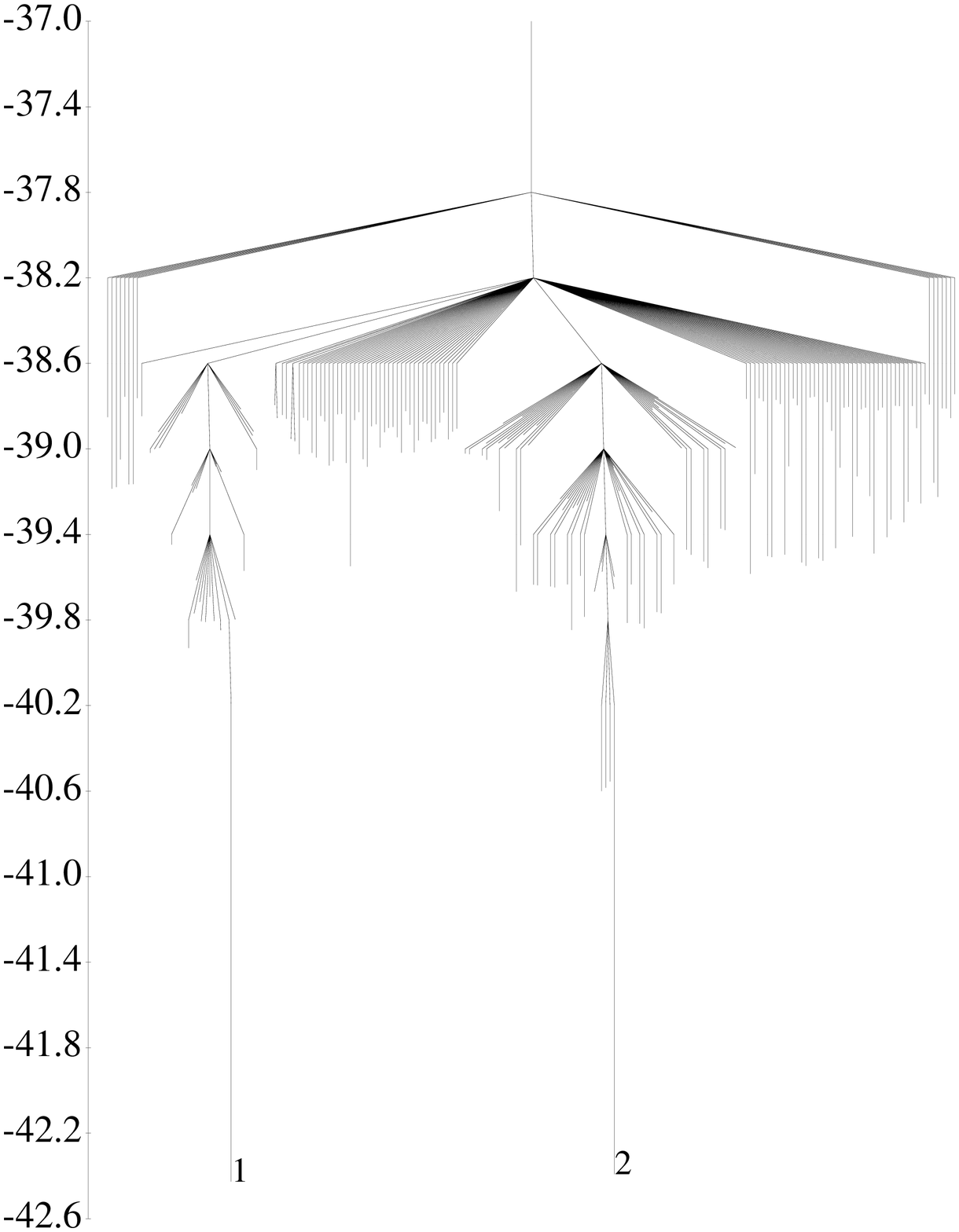} \\
    (a) & (b) \\
    \includegraphics[width=8.5cm,clip=true]{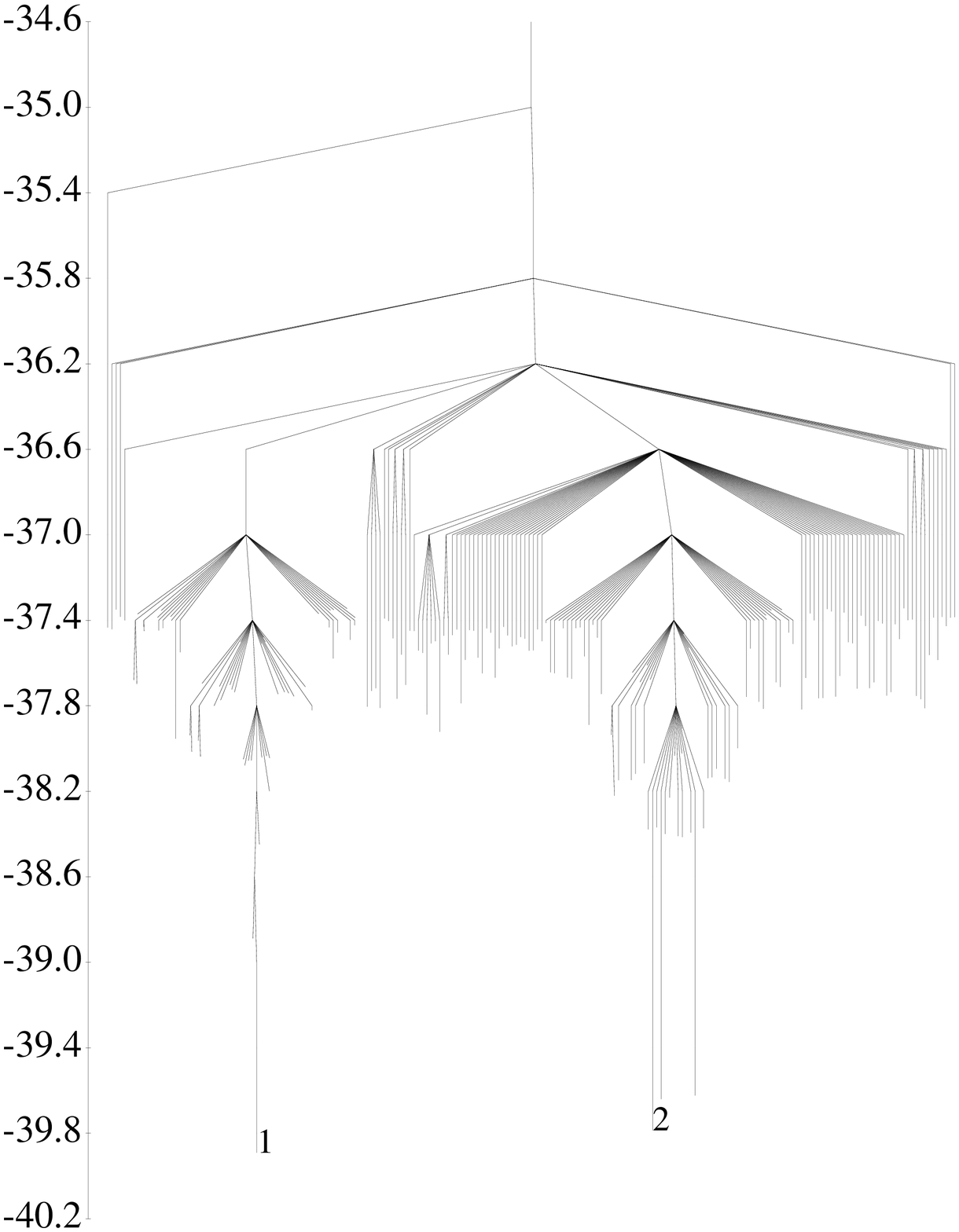} &
    \includegraphics[width=8.5cm,clip=true]{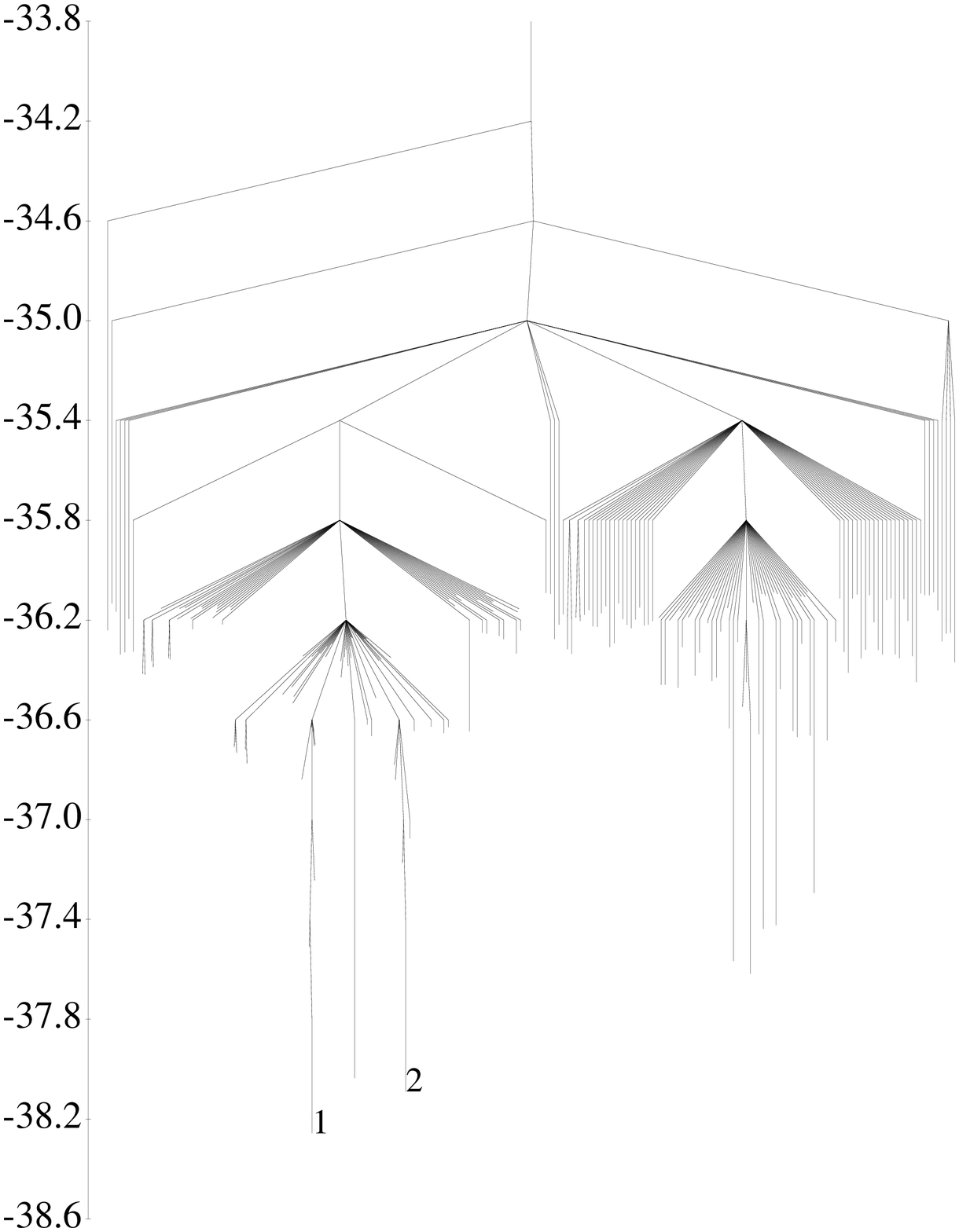} \\
    (c) & (d) \\
        \end{tabular}
\caption{\label{fig:DisCongraph}
Disconnectivity graph for (a) $\mathrm{X_{13}}$, (b) $\mathrm{X_{12}Y_1}$,
(c) $\mathrm{X_{11}Y_2}$, and (d) $\mathrm{X_{10}Y_3}$.
The energy scale is in units of $\epsilon_{XX}$.
The $\mathrm(\sigma,\epsilon)$ values for
panels (a--d) are (1.0,1.0), (0.8,0.6), (0.8,0.5), and (0.8,0.5),
respectively.
Only branches leading to the 200 lowest-energy minima are shown.}
\end{figure*}

\subsection{Quenching procedure}

We have performed minimization or quenching of the parallel tempering Monte 
Carlo (classical and quantum) sampled configurations by implementing 
the Fletcher-Reeves-Polak-Ribiere version of the conjugate gradient method
using the algorithm given in \emph{Numerical Recipes}.\cite{PRESTEU}
The quenches allow us to interpret structural transformations 
in the clusters that are associated with
the peaks in the heat capacity curves as well as variations in the slopes of 
the caloric and the first derivative of the heat capacity curves.

\section{Results and discussion} \label{sec:res}

In this section, we present the results of classical and quantum
parallel tempering Monte Carlo simulations 
performed on each of the $\mathrm{X_{12}Y_1}$,
$\mathrm{X_{11}Y_2}$ and $\mathrm{X_{10}Y_3}$ clusters. The details
about the calculations are given in Section~\ref{sec:comput} .

Figure~\ref{fig:DisCongraph} shows disconnectivity graphs for the studied
system together with a disconnectivity graph for the homogeneous 
$\mathrm{X_{13}}$ cluster. A description of the construction of a
disconnectivity graph is given in the companion paper \cite{SABO04A}
and references therein.
A disconnectivity graph is a useful tool for the visualization
of the underlying potential energy surface of the studied systems.
\cite{KARPLUS97,WALES00}

While the potential energy surface of the homogeneous $\mathrm{X_{13}}$
cluster is characterized by a single funnel,\cite{WALES99C} those 
for $\mathrm{X_{12}Y_1}$, $\mathrm{X_{11}Y_2}$ and $\mathrm{X_{10}Y_3}$
have a double-funnel structure. The global minimum of each system is
labeled by the number 1 and the next higher-lying inherent structure is labeled
by the number 2. For the $\mathrm{X_{12}Y_1}$ [$\mathrm{X_{11}Y_2}$] cluster,
inherent structures 1 ($E=-42.426\epsilon_{XX}$) [($E=-39.891\epsilon_{XX}$)]
and 2 ($E=-42.392\epsilon_{XX}$) [($E=-39.787\epsilon_{XX}$)]
define two basins of similar energies separated
by a large energy barrier. For the $\mathrm{X_{10}Y_3}$ cluster, two
basins are associated with inherent structure 1 ($E=-38.257\epsilon_{XX}$)
(inherent structures 2 and 3 belong to the basin defined by
inherent structure 1) and inherent structure 4 ($E=-37.618\epsilon_{XX}$).
The lowest energy basin of the $\mathrm{X_{12}Y_1}$ cluster contains 30
inherent structures compared with 67 inherent structures in the basin
associates with the second lowest minimum, inherent structure 2. 
Two distinct funnels on the PES of $\mathrm{X_{11}Y_2}$ cluster, one
associated with inherent structure 1 and another with inherent
structure 2, contain 56 and 116 inherent structures, respectively.
Therefore, their lowest energy basin is slightly narrower than
the second basin.
On the other hand, two funnels on the PES of $\mathrm{X_{10}Y_3}$ cluster,
one associated with inherent structure 1 and another with inherent
structure 4, contain 98 and 85 minima, respectively.
The number of inherent structures associated with a basin often indicates
the likelihood that the system will relax to the minimum at the bottom
of that basin.

\begin{figure}[!htbp]
\includegraphics[clip=true,width=8.5cm]{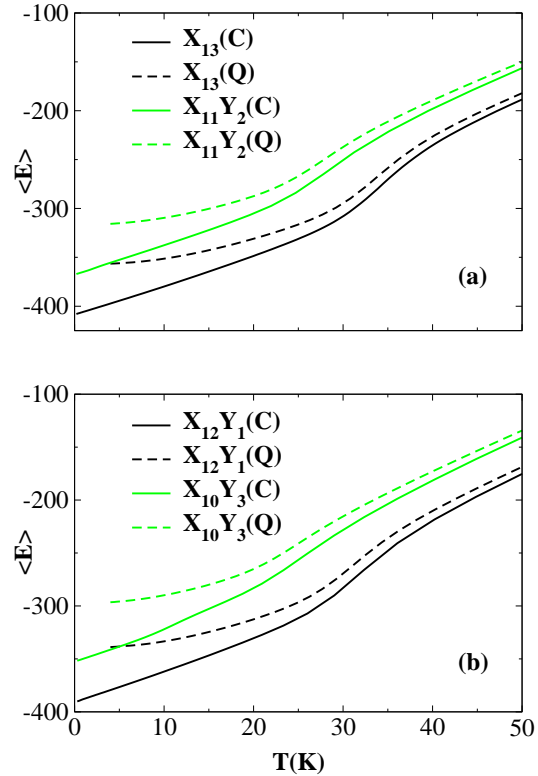}
\caption{\label{fig:avgE} Classical and quantum caloric curves for
the studied systems. Solid (dashed) lines represent classical-C
(quantum-Q) results. $\langle$E$\rangle$ is given in units of Kelvin per
particle. The number of path variables employed in the path integral
simulations is 4$n$=32.
}
\end{figure}
\begin{figure}[!htbp]
\includegraphics[clip=true,width=8.5cm]{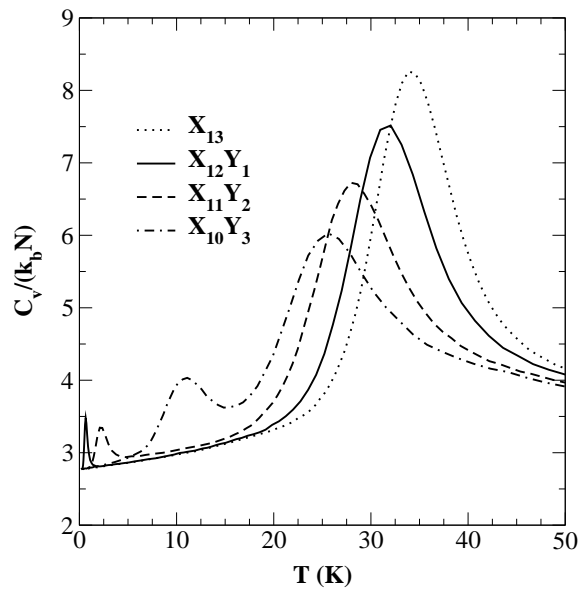}
\caption{\label{fig:cCvall} Classical heat capacities per particle for
studied systems in units of $k_B$ as a function of temperature.
}
\end{figure}
\begin{figure}
\includegraphics[clip=true,width=8.5cm]{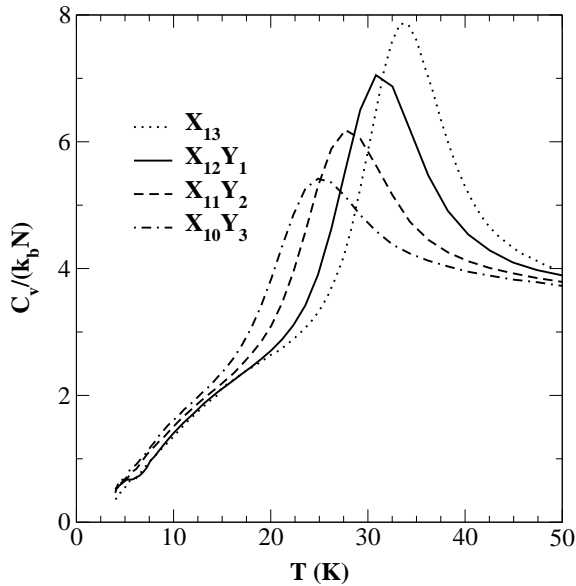}
\caption{\label{fig:qCvall} Quantum heat capacities per particle for
studied systems in units of $k_B$ as a function of temperature. The
number of path variables employed in the simulations is $4n=32$.
}
\end{figure}
Figure~\ref{fig:avgE} displays the caloric curves, i.e. the average total
energy of the clusters, as a function of temperature. Solid lines
represent classical, while dashed lines represent quantum results.
The caloric curves of the homogeneous $\mathrm{X_{13}}$ cluster are
included for comparison purposes. Comparison of the classical and
quantum caloric curves indicates that quantum effects are considerable
over the entire temperature range. A variation in the slope of the
caloric curves is characteristic of a phase change in
the clusters. 

The constant volume heat capacity curves of the studied clusters
as a function of temperature obtained by classical and quantum simulations
are shown in Fig.~\ref{fig:cCvall} and Fig.~\ref{fig:qCvall}, respectively.
Again, the results for the $\mathrm{X_{13}}$ cluster are shown for 
comparison. Replacement of $\mathrm{n}$ $\mathrm{X}$ (Ar) atoms in the 
homogeneous clusters by $\mathrm{n}$ $\mathrm{Y}$ (Ne) atoms 
manifests itself in Fig.~\ref{fig:cCvall} in such a way that the
maxima of the heat capacity curves are shifted toward lower temperatures
with respect to the maximum of the heat capacity of the homogeneous
cluster.
A similar trend is seen for the quantum heat capacities 
(see Fig.~\ref{fig:qCvall}). From Fig.~\ref{fig:cCvall}, it can be
seen that binary clusters exhibit lower temperature peaks, in addition to
higher temperature peaks. Their numerical values and temperatures at which 
they occur are listed in Table~\ref{table:tab.cCv}.
Both higher and lower temperature maxima are to be discussed in more
details in sections below. As can be seen from Fig.~\ref{fig:qCvall},
lower temperature peaks are absent from
the heat capacity curves obtained by path integral simulations, at least
in the temperature range (from 4 K to 50 K) considered here. 
Numerical values of the quantum heat
capacity peaks with associated temperatures are listed in
Table~\ref{table:tab.qCv}.
\begin{table}
\caption{Classical heat capacity peak parameters for 13 atom clusters.
The temperature is given in Kelvin while the heat capacity is given in
units of $k_B$ per particle. The values are obtained by a cubic spline
interpolation of the parallel tempering data shown in Fig.~\ref{fig:cCvall}.
The heat capacity error bars estimates are averages of two
standard deviations of the points near the peaks.}
\label{table:tab.cCv}
\begin{tabular}{lcccc} \hline \hline
 & \multicolumn{2}{c}{Lower temperature peak}
& \multicolumn{2}{c}{Higher temperature peak}  \\ \cline{2-3} \cline{4-5}
           &$T_{peak}$ &$\langle C_V \rangle/k_bN$
&$T_{peak}$ &$\langle C_V \rangle/k_bN$ \\ \hline
X$_{13}$       & &                          & 34.09$\pm$0.18&8.27$\pm$0.06 \\
X$_{12}$Y$_1$  &0.65$\pm$0.01 &3.48$\pm$0.02& 31.67$\pm$0.08&7.54$\pm$0.05 \\
X$_{11}$Y$_2$  &2.21$\pm$0.07 &3.36$\pm$0.02& 28.18$\pm$0.24&6.73$\pm$0.04 \\
X$_{10}$Y$_3$  &11.09$\pm$0.17&4.03$\pm$0.02& 25.61$\pm$0.05&6.03$\pm$0.04 \\
\hline \hline
\end{tabular}
\end{table}
\begin{table}
\caption{Quantum heat capacity peak parameters for 13 atom clusters.
The temperature is given in Kelvin while the heat capacity is given in
units of $k_B$ per particle. The values are obtained by a cubic spline
interpolation of the parallel tempering data shown in Fig.~\ref{fig:qCvall}.
The number of path variables employed in the quantum calculations is
$4n$=32.
The heat capacity error bars estimates are averages of two
standard deviations of the points near the peaks.}
\label{table:tab.qCv}
\begin{tabular}{lcccc} \hline \hline
 & \multicolumn{2}{c}{Lower temperature peak}
& \multicolumn{2}{c}{Higher temperature peak}  \\ \cline{2-3} \cline{4-5}
           &$T_{peak}$ &$\langle C_V \rangle/k_bN$
&$T_{peak}$ &$\langle C_V \rangle/k_bN$ \\ \hline
X$_{13}$       & & & 33.68$\pm$0.21&7.89$\pm$0.04 \\
X$_{12}$Y$_1$  & & & 31.23$\pm$0.27&7.07$\pm$0.04 \\
X$_{11}$Y$_2$  & & & 27.88$\pm$0.30&6.16$\pm$0.04 \\
X$_{10}$Y$_3$  & & & 24.86$\pm$0.12&5.42$\pm$0.03 \\
\hline \hline
\end{tabular}
\end{table}
\begin{figure}
\includegraphics[clip=true,width=8.5cm]{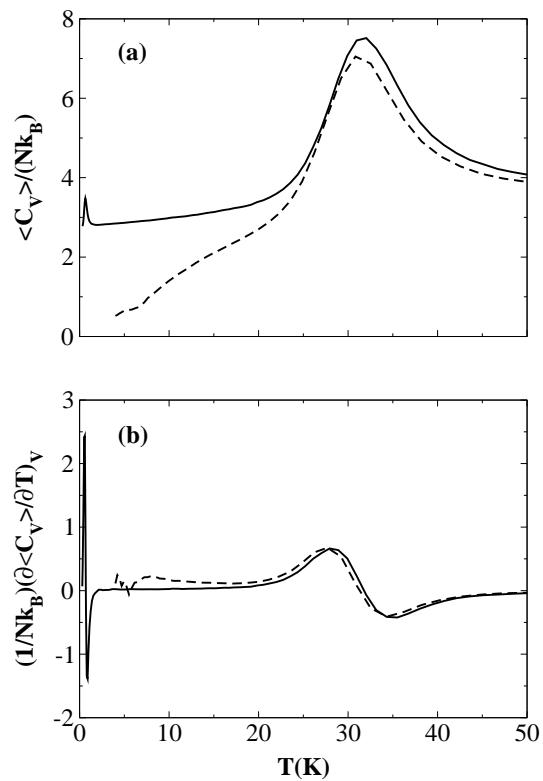}
\caption{\label{fig:dCvdT.x12y1} $\mathrm{X_{12}Y_1}$.
Panel (a) shows classical (solid line) and quantum (dashed line)
heat capacities per particle in units of $k_B$ as a function of temperature.
Panel (b) shows first derivative of the heat capacity per particle in
units of $k_B$. The solid (dashed) line represents classical (quantum) results.
The
number of path variables employed in the path integral simulations is $4n=32$.
}
\end{figure}

\subsection{$\mathrm{X_{12}Y_1}$}

The heat capacity curves and their first derivatives obtained by classical
and path integral Monte Carlo simulations are shown in 
Fig.~\ref{fig:dCvdT.x12y1}(a) and (b), respectively. The solid lines are
classical results.

\subsubsection{Classical simulation}

\begin{figure}
\includegraphics[clip=true,width=8.5cm]{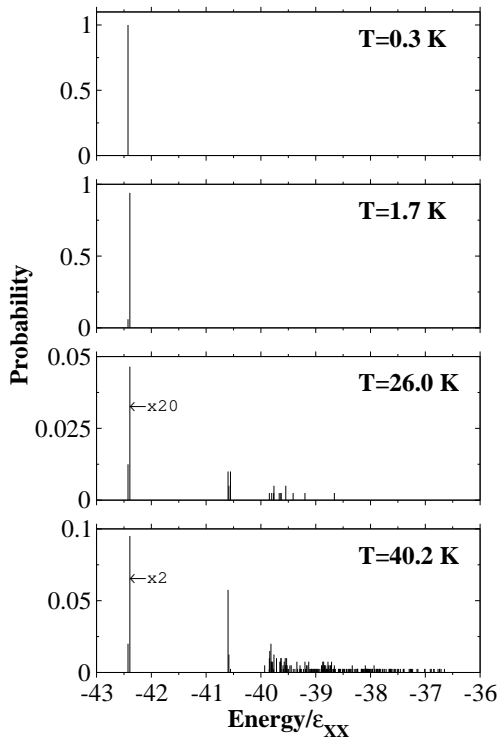}
\caption{\label{fig:prob.c.x12y1} 
Distributions of the inherent structures for the $\mathrm{X_{12}Y_1}$ cluster,
generated by quenching configurations sampled from classical parallel
tempering Monte Carlo simulations at temperatures above and below the
solid $\leftrightarrow$ solid transition (1.7 K and 0.3 K) and above
and below the solid $\leftrightarrow$ liquid transition (40.2 K and 26.0 K).
Note that the lines representing the population of inherent structure 2
at energy -42.392$\epsilon_{XX}$, labeled ``x20'' and ``x2'', have been 
reduced by a factor of twenty and two, respectively.
}
\end{figure}

The heat capacity has a broad peak at a temperature of about 31.7 K and a
narrow, low temperature peak at about 0.65 K. In order to identify the 
phase changes associated with the peaks in the heat capacity curves, the
configurations generated by parallel tempering Monte Carlo simulations
at a temperature somewhat below and above the temperatures of the peaks
in the heat capacity curves have been quenched. The results of the quenches
are shown in Fig.~\ref{fig:prob.c.x12y1}.
Quenching from the configurations sampled at 0.3 K yields only inherent
structure 1 (global minimum-$\mathrm{Y}$ atom is in the interior of the 
cluster; center of the icosahedron). The quenches of the configurations 
sampled at 1.7 K produce predominantly inherent structure 2 ($\mathrm{Y}$ 
atom is on the surface of the cluster), although there
is still a small population (0.06) of inherent structure 1. The low
temperature peak in the classical heat capacity curve is associated with
the structural transition from inherent structure 1 to 
inherent structure 2, that is, a solid $\leftrightarrow$ solid phase change
between two basins. 

The high probability ($\approx0.94$) of finding the $\mathrm{X_{12}Y_1}$ 
cluster to ``dwell'' in inherent structure 2, at a very 
low temperature of 1.7 K, might be surprising at first sight. 
The explanation can be found in the
``shape'' of its disconnectivity graph, see Fig.~\ref{fig:DisCongraph}(b).
As we noticed earlier, by examining its disconnectivity graph,
the energy
landscape displays a double-funnel structure. The funnel associated with
inherent structure 2 contains twice as many local minima than the
funnel associated with inherent structure 1. 
Even though funnel 1 is energetically more favorable,
owing to the larger number of minima associated with funnel 2, the volume
of the configuration space available to funnel 2 is likely to be
larger than one available to funnel 1 and makes it entropicaly more
favorable. 

At a temperature $T=26$ K (below the $T_{peak}$) the system occupies mainly
($\approx0.93$) inherent structure 2. 
However, a very small population of 
inherent structure 1 and higher energy amorphous structures, 
is present. As the temperature increases, the population of higher energy
amorphous
structures rapidly increases and at $T=40.2$~K they begin to dominate.
Thus, the broad peak in the heat capacity at $T=31.67$ K is associated
with the solid $\leftrightarrow$ liquid phase change.

\begin{figure}
\includegraphics[clip=true,width=8.5cm]{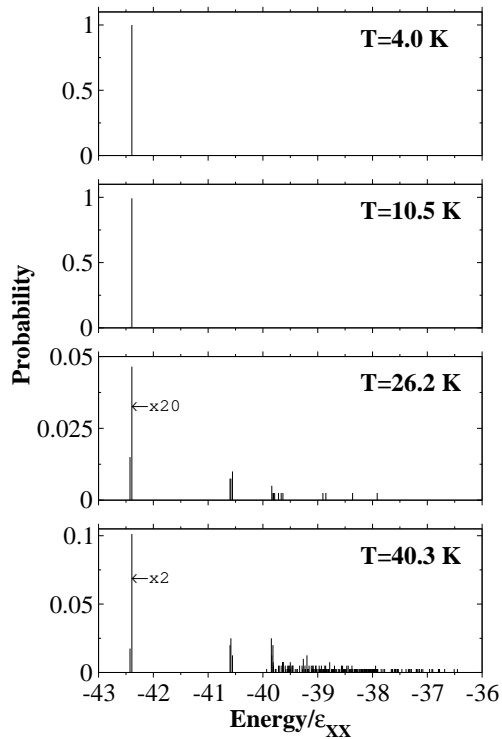}
\caption{\label{fig:prob.q.x12y1} 
Distributions of the inherent structures for the $\mathrm{X_{12}Y_1}$ cluster,
generated by quenching configurations sampled from quantum parallel
tempering Monte Carlo simulations at temperatures 4 K, 10.5 K and
temperatures below (26.2 K) and above (40.3 K) the solid $\leftrightarrow$
liquid transition. Note that the lines representing the population of 
inherent structure 2 at energy -42.392$\epsilon_{XX}$, 
labeled ``x20'' and ``x2'',
have been reduced by a factor of twenty and two, respectively.
}
\end{figure}

\subsubsection{Path integral simulation}

Unlike the classical heat capacity, the quantum heat capacity curve has only
one peak, centered near $T=31.23$~K. It can be seen from 
Fig.~\ref{fig:dCvdT.x12y1}
and by comparing the numerical results from Tables~\ref{table:tab.cCv} and
\ref{table:tab.qCv}, 
the solid-liquid phase change in the ``quantum'' system occurs at slightly 
lower temperature than in the ``classical'' one. Quantum effects lower the
transition temperature by 1.4\%. Likewise, there is a decrease in the height
of the heat capacity maximum. The lowering of the transition temperature
and the height of the heat capacity peak from quantum effects are expected
and have been documented in the literature.\cite{FREEMAN92,CHAKRA95A,
FREEMAN00C,LOPEZ02} 
These phenomena are the consequence of the zero-point motion and/or 
tunneling that effectively raise the energy of the inherent structures, lower 
the energy barriers between them and thus allow for easier isomerization.

From the distribution of inherent structures at $T=4$~K it can
be seen that the system ``likes'' to stay in inherent structure 2 with
the probability 1.0 . A normal mode analysis of inherent structures 1 and
2 shows that estimated zero-point energy of inherent structure 1 lies
about 22 cm$^{-1}$ higher than the corresponding zero-point energy associated
with inherent structure 2.
At the temperature of 10.5 K, there is still a very
high population ($\approx 0.99$) of inherent structure 2 with almost 
a negligible population of inherent structure 1 ($\approx 0.01$). 
We have compared the classical distribution of inherent structures at 
approximately the same temperatures ($T=4.09$ K and $T=10.9$~K) with its
quantum counterpart. It has been found that the global minimum in the 
classical case is slightly more likely to be populated than in the quantum 
case. At the temperatures below (26.2 K) and above (40.3 K) the temperature
at which the solid $\leftrightarrow$ liquid phase change occurs, the 
classical and quantum distributions of inherent structures are similar
\begin{figure}
\includegraphics[clip=true,width=8.5cm]{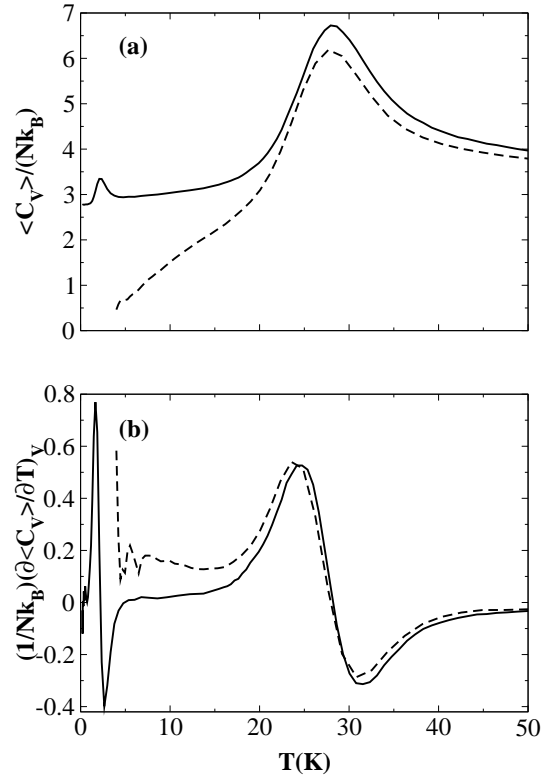}
\caption{\label{fig:dCvdT.x11y2} $\mathrm{X_{11}Y_2}$.
Panel (a) shows classical (solid line) and quantum (dashed line)
heat capacities per particle in units of $k_B$ as a function of temperature.
Panel (b) shows first derivative of the heat capacity per particle in
units of $k_B$. The solid (dashed) line represents classical (quantum) results.
The
number of path variables employed in the path integral simulations is $4n=32$.
}
\end{figure}

\subsection{$\mathrm{X_{11}Y_2}$}

The heat capacity curves and their first derivatives obtained by classical
and path integral Monte Carlo simulations are shown in
Fig.~\ref{fig:dCvdT.x11y2}(a) and (b), respectively. The solid lines are
classical results.

\subsubsection{Classical simulation}

As is the case with $\mathrm{X_{12}Y_1}$, the heat capacity curve of  
the $\mathrm{X_{11}Y_2}$ cluster has two peaks.
A broad, high temperature peak occurs at 28.18 K and a smaller, narrower
peak occurs at 2.21 K. Quenching of the configurations generated at $T=1.2$ K
gives only inherent structure 1 (one $\mathrm{Y}$ atom is in the interior and
the other is on the surface of the cluster). The quenches of the 
configurations generated at $T=4.2$ K yield inherent structures
1, 2 and 3 with the population probability of approximately 0.2, 0.77 
and 0.03, respectively. Inherent structures 2 and 3 belong to the same
funnel. Thus, the low temperature peak in the heat capacity curve is
associated with the structural transition between inherent structure 1
and inherent structure 2 and correspond to a solid $\leftrightarrow$
solid phase change.
\begin{figure}
\includegraphics[clip=true,width=8.5cm]{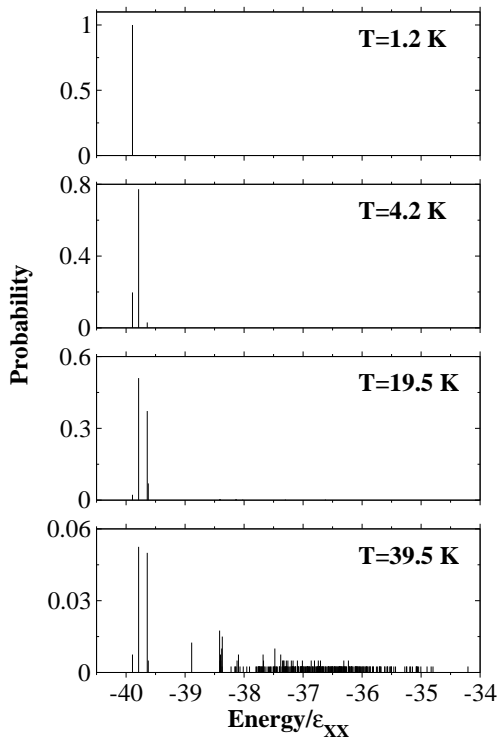}
\caption{\label{fig:prob.c.x11y2} 
Distributions of the inherent structures for the $\mathrm{X_{11}Y_2}$ cluster,
generated by quenching configurations sampled from classical parallel
tempering Monte Carlo simulations at temperatures above and below the
solid $\leftrightarrow$ solid transition (4.2 K and 1.2 K) and above
and below the solid $\leftrightarrow$ liquid transition (39.5 K and 19.5 K).
}
\end{figure}

The high probability of occupying inherent structure 2 at a relatively
low temperature can be explained in an analogous manner to that
for the $\mathrm{X_{12}Y_1}$ cluster.
The funnel associated with inherent structure 2 [see
Fig.~\ref{fig:DisCongraph}(b)] contains two times more minima than the one
associated with inherent structure 1 and has a larger entropy. 

At a temperature $T=19.5$ K, inherent structures 2, 3 and 4 associated
with funnel 2 dominate in the quench distribution. There is also
a very small population of the amorphous (glassy) structures.
At $T=39.5$ K, which is on the high temperature side of the higher peak
in the heat capacity curve, the inherent structures from the basin 2
still dominate, but there is an appreciable population of the higher energy
amorphous structures. This indicates that the cluster undergoes a phase
change from a solid-like to a liquid-like form.
\begin{figure}
\includegraphics[clip=true,width=8.5cm]{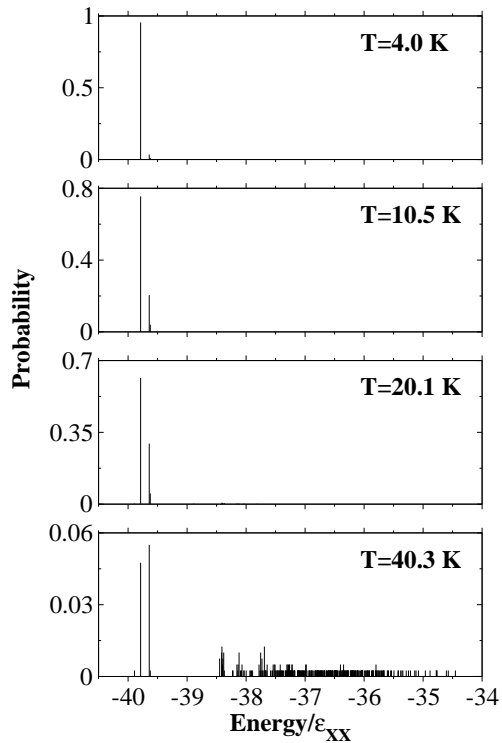}
\caption{\label{fig:prob.q.x11y2} 
Distributions of the inherent structures for the $\mathrm{X_{11}Y_2}$ cluster,
generated by quenching configurations sampled from quantum parallel
tempering Monte Carlo simulations at temperatures 4 K, 10.5 K and
temperatures below (20.1 K) and above (40.3 K) the solid $\leftrightarrow$
liquid transition. 
}
\end{figure}

\subsubsection{Path integral simulation}

In the temperature range examined in this work,
the quantum heat capacity curve has only one peak at $T=27.88$ K. From
Fig.~\ref{fig:dCvdT.x11y2} one can see that the quantum effects
affect the phase change in the $\mathrm{X_{11}Y_2}$ cluster by shifting
the solid-liquid transition temperature 
toward lower temperatures and decreasing the
hight of the maximum of the heat capacity curve. The change in the
transition temperature with respect to the classical result is
approximately 1.1\%.

The quenches from the configurations sampled by quantum parallel tempering 
Monte Carlo simulations at 4 K yield inherent structures 2, 3 and 4 (associated
with the basin 2) with the probability 0.95, 0.04 and 0.01, respectively.
There is no measurable population of the global minimum at that temperature.
Comparing the quantum distribution of the inherent structures at 4 K
with the classical distribution at approximately the same temperature,
$T=4.2$ K, we see that there is a sizeable population of the global
minimum in the classical results. Why is that?
An explanation is in the quantum effects, i.e. zero-point energy.
We have performed a normal mode analysis of inherent structures 1 and
2 and have found that the estimate of zero-point energy of inherent
structure 1 lies about 30 cm$^{-1}$ higher than the corresponding zero-point
energy associated with inherent structure 2.
At $T=10.5$ K, the quenching again yields inherent structures 2, 3 and 4
but with probability 0.75, 0.21 and 0.04, respectively.
At the temperatures below (20.1 K) and above (40.3 K) the temperature
at which the solid $\leftrightarrow$ liquid phase change occurs, the
classical and quantum distributions of the inherent structures are similar
\begin{figure}
\includegraphics[clip=true,width=8.5cm]{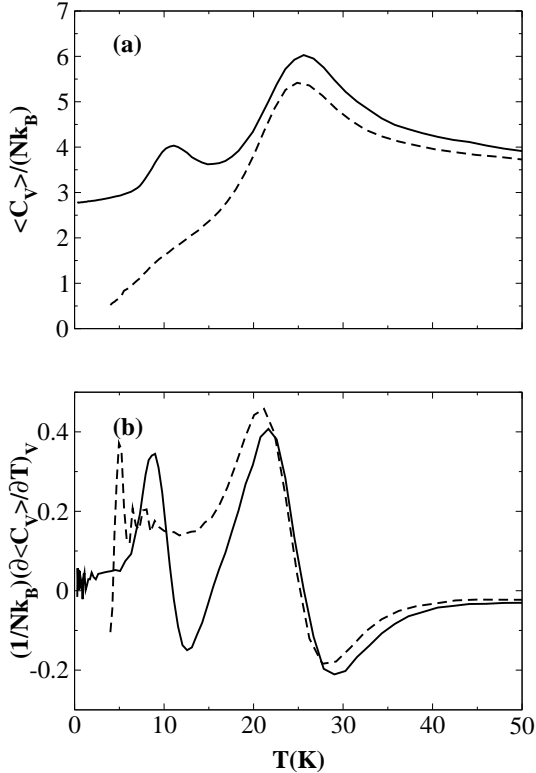}
\caption{\label{fig:dCvdT.x10y3} $\mathrm{X_{10}Y_3}$.
Panel (a) shows classical (solid line) and quantum (dashed line)
heat capacities per particle in units of $k_B$ as a function of temperature.
Panel (b) shows first derivative of the heat capacity per particle in
units of $k_B$. The solid (dashed) line represents classical (quantum) results.
The
number of path variables employed in the path integral simulations is $4n=32$.
}
\end{figure}

\subsection{$\mathrm{X_{10}Y_3}$}

The heat capacity curves and their first derivatives obtained by classical
and path integral Monte Carlo simulations are shown in
Fig.~\ref{fig:dCvdT.x10y3}(a) and (b), respectively. The solid lines are
classical results.

\subsubsection{Classical simulation}

The heat capacity has a broad peak at a temperature of about 25.6 K and a
narrower, low temperature peak at about 11.1 K. Quenching of the 
configurations generated at 5.1 K obtains inherent structures 1 and 2
with probabilities of 0.99 and 0.01, respectively.
Both inherent structures 1 and 2 belong to the funnel associated 
with inherent structure 1.
At the temperature 15.1 K, the quenches of the sampled configurations
obtain two groups of inherent structures. One group, comprised of the low-lying
inherent structures (1, 2, 3 and 6), belongs to the funnel associated
with inherent structure 1 and other group, comprised of the low-lying
inherent structures (4, 5, 7, 8 and 9) that belongs to the funnel associated
with inherent structure 4. The population probability of the low-lying
inherent structures at the bottom of the funnel associated with inherent
structure 1 is 0.22 while for those that belong to the funnel associated
with inherent structure 4 is 0.77. There is also a very small population
of the higher-energy, glassy-like structures.
The low temperature heat capacity peak can be characterized as a phase
change arising from the structural transition between two groups of the
inherent structures rather than the structural transition between 
individual inherent structures, which is the case for the $\mathrm{X_{12}Y_1}$
and $\mathrm{X_{11}Y_2}$ clusters. Therefore, it is characterized as 
a solid $\leftrightarrow$ solid phase change.
\begin{figure}
\includegraphics[clip=true,width=8.5cm]{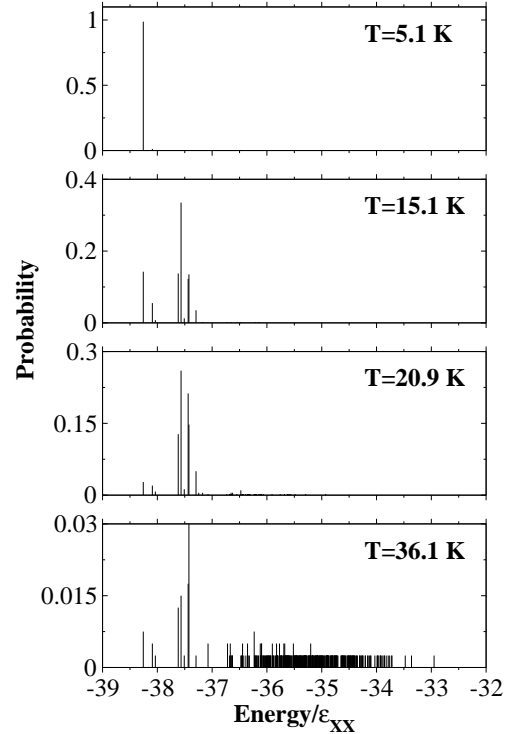}
\caption{\label{fig:prob.c.x10y3} 
Distributions of the inherent structures for the $\mathrm{X_{10}Y_3}$ cluster,
generated by quenching configurations sampled from classical parallel
tempering Monte Carlo simulations at temperatures above and below the
solid $\leftrightarrow$ solid transition (15.1 K and 5.1 K) and above
and below the solid $\leftrightarrow$ liquid transition (36.1 K and 20.9 K).
}
\end{figure}
As noticed earlier, the energy landscape of $\mathrm{X_{10}Y_3}$ has
two funnels with approximately the same number of associated inherent
structures. 
As a consequence, it is equally likely that the system upon cooling 
enters either of the funnels. At the relatively low temperature of 5.1 K
the system dwells in the funnel associated with the global minimum,
which is not the case for the $\mathrm{X_{12}Y_1}$ and $\mathrm{X_{11}Y_2}$
clusters as discussed earlier.

At $T=20.9$ K, the group of inherent structures that belongs to the funnel
associated with inherent structure 4 dominates in the quench distribution.
However, there is a large number of amorphous structures that start
to be populated. At $T=36.1$ K the population of amorphous structures
rapidly increases. The heat capacity peak at $T=25.61$ K is, therefore, 
associated with the solid $\leftrightarrow$ liquid phase change.
\begin{figure}
\includegraphics[clip=true,width=8.5cm]{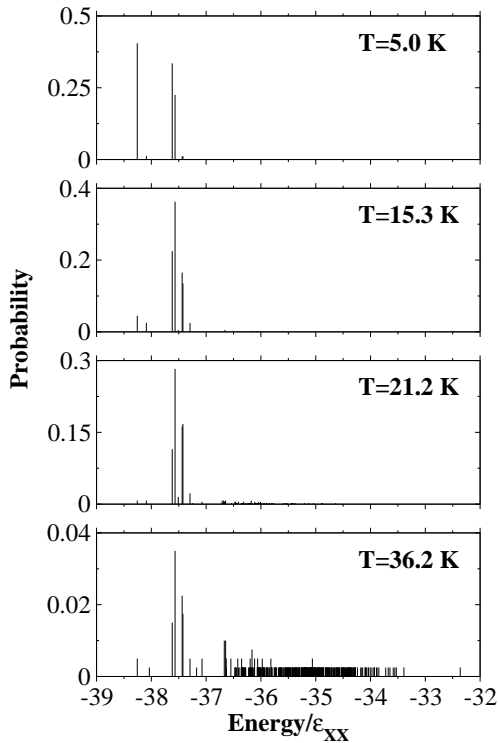}
\caption{\label{fig:prob.q.x10y3} 
Distributions of the inherent structures for the $\mathrm{X_{10}Y_3}$ cluster,
generated by quenching configurations sampled from quantum parallel
tempering Monte Carlo simulations at temperatures 5 K, 15.3 K and
temperatures below (21.2 K) and above (36.2 K) the solid $\leftrightarrow$
liquid transition.
}
\end{figure}

\subsubsection{Path integral simulation}

Unlike the classical heat capacity, in the range of temperatures examined
in the current work, the quantum heat capacity curve has
only one peak, at the temperature of 24.86 K. This single peak behavior can 
be seen from Fig.~\ref{fig:dCvdT.x10y3} and by comparing the numerical results
from Tables~\ref{table:tab.cCv} and \ref{table:tab.qCv} that the quantum
effects lower the transition temperature by 2.9\% with respect to the
classical result.

The quenches from the configurations sampled at 5.0~K yield two groups
of inherent structures with approximately the same probability. 
Group 1 contains inherent structures 1 and 2, and they belong to the
funnel associated with inherent structure 1. Group 2 is made up
of inherent structures 4, 5, 7 and 8, and they belong to the funnel
associated with inherent structure 4. Comparing the quantum distribution
of the inherent structures with its classical counterpart 
at the same temperature, one can see that quantum effects
allow for an easier isomerization between two funnels. Moreover,
the quantum effects eliminate the low temperature peak in the quantum heat
capacity curve over the studied temperature range. 
At $T=15.3$ K two groups of inherent structures, group 1 and 2,
dominate in the quench distribution. When compared to the classical
distribution of inherent structures at approximately the same temperature,
one can see that the population probability of group 1 is smaller than
in the classical case. At the higher temperatures, the classical and quantum
distributions of the inherent structures become similar and amorphous
structures dominate. As expected, the heat capacity peak at $T=24.86$ K
is associated with the solid $\leftrightarrow$ liquid phase change. 

\section{Conclusions} \label{sec:conclude}

In this work, we have investigated the melting behavior of the selected
binary Lennard-Jones clusters of the form $\mathrm{X_{13-n}Y_n}$ in both
classical and quantum regimes. For the classical results, the total energy, 
the heat capacity and the first derivative of the heat capacity, are obtained 
by employing classical Monte Carlo method in conjunction with parallel
tempering, while their quantum counterparts are obtained by employing 
the path integral (RW-WFPI) Monte Carlo method combined with parallel
tempering. Quenching of the Monte Carlo sampled configurations permits
us to identify the structural transitions associated with the peaks
in the heat capacity curves.

Classical results show that all three studied systems exhibit a low
temperature peak in the heat capacity curve. These low temperature
peaks result from a structural transformation between two solid-like low
lying, close in energy, inherent structures or groups of inherent structures.
In other words, low temperature peaks are a direct consequence of the 
double-funnel structure of the underlying potential energy surface of 
the system. The high temperature peaks are associated with a solid to liquid 
melting transition.

Replacing $\mathrm{n}$ $\mathrm{X}$ (core) atoms in the homogeneous cluster  
with $\mathrm{n}$ lighter impurity atoms $\mathrm{Y}$
results in the lowering of the melting temperature and
the hight of the peak of the classical heat capacity when compared to the same
quantities in the homogeneous cluster (see Fig.~\ref{fig:cCvall}). The
quantum heat capacity shows a similar trend (see Fig.~\ref{fig:qCvall}).
On the other hand, low temperature peaks in the classical heat capacity curves 
are shifted toward higher temperatures as the number of atoms $\mathrm{Y}$
increases in the clusters.

Quantum effects are important but relatively modest in the whole range 
of temperatures except in the low temperature regime where
they are more pronounced. This is understandable because the dominant
species in the clusters are $\mathrm{X}$ atoms that mimic (heavier)
Ar atoms. As the number of the $\mathrm{Y}$ atoms, that mimic
(lighter) Ne atoms, increases so does the magnitude of the quantum effect.
In general, the magnitude of the quantum effects depends on the particles
mass, the temperature and the relative strengths of the like and mixed pair
interaction potential.
Quantum effects do lower the melting temperatures in all three studied
systems. Moreover, the low temperature peaks present in the classical
heat capacities of the clusters disappear in the quantum case (at least 
in the temperature range considered in this work).

{\bf Acknowledgment}

The authors acknowledge support from the National Science Foundation
through awards No. CHE-0095053. and CHE-0131114. One of us (D.S.) would like
to thank Cristian Diaconu for helpful discussions. The authors would
also like to thank Brown University's Center for Advanced Scientific
Computing and Visualization for their assistance with the present research.

\bibliography{/aux/sabo/revtex4/revtex4/papers/sabo}


\end{document}